\documentclass[sigconf]{acmart}
\usepackage{libertine}
\usepackage{xspace}

\usepackage{balance}       
\usepackage{graphics}      
\usepackage[T1]{fontenc}   
\usepackage{color}
\usepackage{booktabs}
\usepackage{textcomp}
\usepackage{multirow}

\usepackage{CJKutf8}

\usepackage{microtype}        
\usepackage{ccicons}          

\newcommand{\tool}{{\texttt{GptVoiceTasker}}\xspace}
\newcommand{\chen}[1]{\textcolor{red}{\textbf{Chen}: #1}}
\newcommand{\uist}[1]{\textcolor{black}{#1}}

\def\pprw{8.5in}
\def\pprh{11in}

\setcopyright{acmlicensed}
\acmConference[UIST'24]{ACM Symposium on User Interface Software and Technology}{October 13--16, 2024}{Pittsburgh, PA, USA} \acmDOI{10.1145/3654777.3676356}

\begin{document}

\title{GPTVoiceTasker: Advancing Multi-step Mobile Task Efficiency Through Dynamic Interface Exploration and Learning}

\author{Minh Duc Vu}
\email{dustin.vu@csiro.au}
\orcid{https://orcid.org/0000-0002-4798-8701}
\affiliation{%
  \institution{CSIRO's Data61}
  \city{Melbourne}
  \country{Australia}}
\author{Han Wang}
\email{han.wang@monash.edu}
\orcid{https://orcid.org/0000-0001-7862-6677}
\affiliation{%
  \institution{Monash University}
  \city{Melbourne}
  \country{Australia}}
\authornote{Minh Duc Vu and Han Wang contributed equally.}
  
\author{Zhuang Li}
\email{zhuang.li@monash.edu}
\orcid{https://orcid.org/0000-0002-9808-9992}
\affiliation{%
  \institution{Monash University}
  \city{Melbourne}
  \country{Australia}}

\author{Jieshan Chen}
\email{jieshan.chen@data61.csiro.au}
\orcid{https://orcid.org/0000-0002-2700-7478}
\affiliation{%
  \institution{CSIRO's Data61}
  \city{Sydney}
  \country{Australia}}

\author{Shengdong Zhao}
  \affiliation{%
  \institution{City University of Hong Kong}
  \city{Hong Kong}
  \country{China}}
\email{shezhao@cityu.edu.hk}
\orcid{https://orcid.org/0000-0001-7971-3107}

\author{Zhenchang Xing}
  \affiliation{%
  \institution{CSIRO's Data61 \& Australian National University}
  \city{Canberra}
  \country{Australia}}
\email{zhenchang.xing@data61.csiro.au}
\orcid{https://orcid.org/0000-0001-7663-1421}

\author{Chunyang Chen}
\affiliation{%
  \institution{Technical University of Munich \& Monash University}
  \city{Heilbronn}
  \country{Germany}}
\email{chun-yang.chen@tum.de}
\orcid{https://orcid.org/0000-0003-2011-9618}
\authornote{Chunyang Chen is the corresponding author.}

\renewcommand{\shortauthors}{Vu et al.}

\begin{abstract}
  
Virtual assistants have the potential to play an important role in helping users achieves different tasks. However, these systems face challenges in their real-world usability, characterized by inefficiency and struggles in grasping user intentions. Leveraging recent advances in Large Language Models (LLMs), we introduce \tool, a virtual assistant poised to enhance user experiences and task efficiency on mobile devices. \tool excels at intelligently deciphering user commands and executing relevant device interactions to streamline task completion. For unprecedented tasks, \tool utilises the contextual information and on-screen content to continuously explore and execute the tasks. In addition, the system continually learns from historical user commands to automate subsequent task invocations, further enhancing execution efficiency. \uist{From our experiments, \tool achieved 84.5\% accuracy in parsing human commands into executable actions and 85.7\% accuracy in automating multi-step tasks.} In our user study, \tool boosted task efficiency in real-world scenarios by 34.85\%, accompanied by positive participant feedback. We made \tool open-source, inviting further research into LLMs utilization for diverse tasks through prompt engineering and leveraging user usage data to improve efficiency.
\end{abstract}


\begin{teaserfigure}
\centering
\includegraphics[width=\textwidth]{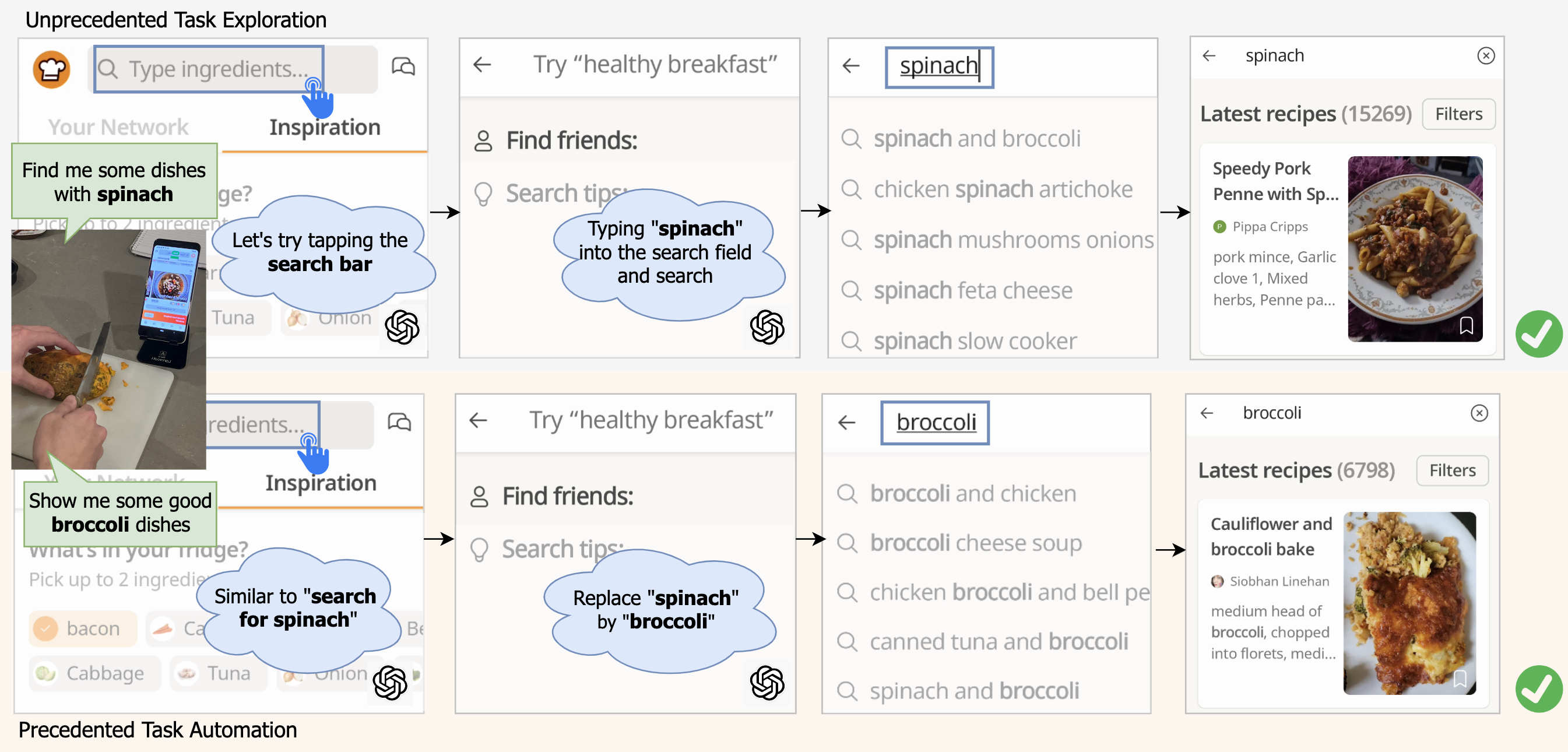}
\caption{\tool provides an intuitive method for automating complex commands on smartphones during physically demanding activities, such as cooking. It automatically explores step-by-step interactions to complete unprecedented tasks and uses the saved information to accelerate the automation process for tasks that have been previously encountered.}
\label{fig:thumbnail_image}
\end{teaserfigure}

\maketitle

\footnotetext{This paper has been accepted and will be presented at UIST 24.}


\section{Introduction}
\label{sec:introduction}

The advancements in voice control technology have sparked a new wave of innovation, driving the exploration of its potential in transforming smartphone interactions~\cite{hoy2018alexa,hirschberg_manning_2015}. With the integration of voice control, users can effortlessly navigate through various applications, compose messages, and even initiate tasks like checking the weather or playing a video on YouTube~\cite{voicify}. This natural mode of interaction not only saves time but also promotes a hands-free experience, allowing individuals to engage with their smartphones in situations where manual input operations are impractical or inconvenient~\cite{zhong_raman_burkhardt_biadsy_bigham_2014,liu2015voice} (refer to Figure.~\ref{fig:overview} illustrating a user engaged in gym activities). Moreover, the success stories of widely recognized voice assistants like Google Voice Assistant~\cite{google_assistant} and Siri~\cite{siri} have further propelled the adoption of voice-based interactions, inspiring researchers and developers to delve deeper into its capabilities and refine its usability for an even broader range of users. 

Developing autonomous voice-controlled assistants involves addressing various challenges that impact the usability of these systems~\cite{voice_problems}. 
Current industrial products (such as Siri and Google Assistant) do not provide a universal approach as they require app developers to explicitly define a narrow set of voice-supported actions within the code. For instance, while those voice assistants fully support searching for videos on YouTube, other YouTube features such as accessing video history is unavailable. This results in a disjointed and often unfinished experience for users as they cannot rely on voice interactions to fully control their smartphone. In addition, these assistants often struggle to comprehend commands when users' inputs are unclear or do not align with the predefined patterns~\cite{custom_intent_2024, apple_intent_2024}. This lack of understanding further impacts the overall utility of the voice-controlled systems. Furthermore, app developers face the challenge of anticipating and programming an extensive variety of potential intents, a task that is both complex and limiting in the scope of voice assistant capabilities. 

Recently, Large Language Models (LLMs) have brought a paradigm shift in natural language processing (NLP), demonstrating remarkable capabilities in tasks like reading comprehension, translation, and text completion~\cite{bareiss2022code}. The advent of Few-Shot Learning has further amplified the capabilities of LLMs, enabling them to quickly adapt to new logical reasoning tasks with minimal examples. This versatility and efficiency in managing various conversational interactions without the need for extensive retraining offer a groundbreaking approach, obviating the need for task-specific models and extensive datasets~\cite{perez2021true}.
In the context of mobile assistants/agents, the integration of LLMs to comprehend user commands and interface with mobile UIs has gained traction~\cite{wen2023empowering, wang2023enabling, yan2023gpt, pan2023autotask, carreira2023revolutionizing}. Amazon's integration of LLMs into Alexa represents a significant step in enhancing voice assistants~\cite{alexa_llm}. Their primary focus has been on improving Alexa's ability to understand users' needs more accurately and to control other devices more effectively. But it's not clear how Alexa can help with smartphones users and if the assistant is optimized with mobile UIs.  Some research in this area concentrates on translating mobile GUIs into text, relying on LLMs to understand the context and predict interactive screen elements~\cite{wang2023enabling, wen2023empowering}. However, this method sometimes struggles with the extraction of irrelevant GUI elements or fails when the command does not directly relate to the current screen. Other UI automation tools for UI testing explored enhancing LLMs with image processing capabilities, such as those found in GPT-4v~\cite{yan2023gpt}. While this approach shows a higher success rate, it is hampered by longer processing times and increased costs, which can negatively impact user experience in real-time systems like voice assistants.

This paper introduces \tool, a novel voice assistant that automates multi-step unprecedented tasks by dynamically exploring app interface and accelerate similar tasks through prior usages. Drawing inspiration from the conventional record-and-replay approach~\cite{lam2017record, li2017sugilite}, \tool is designed to learn GUI transformations as users navigate through apps. Instead of simply transmitting live UI information, \tool captures information of the current UI as the user interacts with the app, storing this data in a backend database as known knowledge to the system. Whenever a user command is issued, \tool cross-references the new executing task with this stored knowledge to make informed decisions. This method allows reproduction of tasks for similar future requests, enhancing task efficiency and accuracy. If a user command references a unprecedented feature, \tool will step-by-step explore and record the navigational path to achieve the feature. 
Our system achieves this by advanced prompt engineering techniques to ensure a precise understanding of user commands without extensive model training. 
\tool effectively bridges the divide between natural language commands and interactive mobile tasks, enabling seamless automation of everyday tasks that include actions like scrolling, tapping, and text input purely via voice commands.

We validated the technical contributions of \tool by evaluating i) the ability to parse user commands into executable actions, ii) the ability to complete multi-step tasks given one command, and iii) the ability to streamline saved tasks. The command parser achieved over 90\% accuracy on a human command dataset collected from the user study. \tool also achieved 85.7\% success rate in completing human-collected multi-step tasks. Our automated execution achieved 82.7\% success rate for direct match tasks and 72.0\% success rate for tasks with different parameters. To validate the usability of \tool, we conducted a user evaluation with 18 participants, each completing a set of tasks using \tool and two state-of-the-art baselines. We collected the time taken to complete each task, as well as quantitative and quality feedback from users. The results showed that \tool accelerated the tasks by 34.85\% and received positive feedback regarding usability.

To summarize, the contributions of this paper include:
\begin{itemize}

    \item Development of \tool, a voice assistant that harnesses the capabilities of LLMs to streamline the automation of multi-step tasks by predicting the most optimal step on each individual screen.
    \item A graph-based local database design that automates the recording and retrieval of personal app usages, enhancing task execution efficiency for virtual assistant interactions.
    \item Conducting a user evaluation to validate the effectiveness of our approach, along with empirical findings on system limitations and considerations for voice assistant design.
    \item \tool\footnote{\url{https://github.com/vuminhduc796/GPTVoiceTasker}} is open-sourced so that anyone can use and continue to improve the system. 
\end{itemize}

\section{Background \& Related works}
\label{sec:relatedWorks}


\subsection{Voice Control \& Automation on Mobile Devices}

Recent advancements in Natural Language Understanding (NLU) have significantly enhanced the development of voice assistants across various platforms, including ubiquitous systems~\cite{krishna2012zigbee,bai2022research} and home appliances~\cite{park2018low}. 

An early milestone in smartphone voice control interfaces was JustSpeak, which harnessed Google's Automatic Speech Recognition (ASR) to record user commands and introduced innovative utterance parsing techniques~\cite{zhong_raman_burkhardt_biadsy_bigham_2014}. Subsequently, the Smart Voice Assistant expanded on JustSpeak's capabilities by enabling users to manage calls and SMS through voice commands~\cite{bhalerao2017smart}. However, these initial approaches, foundational as they were, encountered usability issues stemming from rigid language parsing heuristics and limited use cases, which spurred the need for further development of smartphone virtual assistants.

In recent years, significant advancements in language parsing capabilities have been achieved through deep learning models. SAVANT leveraged Dialogflow as a conversational agent to extract user intent from utterances~\cite{arsan2021app}, while DoThisHere employed the pre-built Almond language model to enable voice control for retrieving and setting UI contents in Android~\cite{yang2020dothishere}. Google released Voice Access~\cite{voice_access_issue}, aimed to replace manual interactions with voice command, which has over 100 millions downloads on Google Play Store. \uist{ Moreover, there are Firefox Voice, an open and extensible web-based voice assistant with speech-to-text engine~\cite{firefox2021chi}, and Talk2Care, which leverages LLMs to facilitate communication between healthcare providers and older adults~\cite{talk2care2024imwut}. Additionally, Just Speak It has focused on minimizing cognitive load during eyes-free text editing with a smart voice assistant~\cite{justspeakit2021uist}, while GazePointAR uses context-aware multimodal inputs for pronoun disambiguation in augmented reality~\cite{gazepointar2024chi}.} Voicify~\cite{voicify} introduced VoicifyParser, an advanced deep learning approach for parsing user commands into on-screen interactions. However, the interaction paradigm with these existing approaches remains somewhat unnatural, requiring users to issue precise machine-like instructions, such as ``\emph{Press save button}''. This limitation means that they may struggle to fully comprehend high-level user intentions, such as ``\emph{I want to save this note}''. We propose \tool to address these challenges and revolutionise the voice-based interactions between human and software systems. Our solution leverages the capabilities of LLMs to map high-level user intentions to executable actions, enabling on-screen interactions through intention-based voice commands. This approach seeks to accommodate the flexibility and natural language of human commands, ushering in a new era of user-friendly assistive tools.

Research has also delved into voice command interfaces for automating smartphone tasks, often categorized as programming-by-demonstration tools~\cite{arsan2021app}. These systems generally utilize a record-and-replay strategy, where the user records a series of actions to complete a task and later triggers that sequence with a voice command. SUGILITE~\cite{li2017sugilite}, for instance, introduced methods for performing task variations with different parameters from a single recorded instance. Building upon this, AutoVCI~\cite{autovci} automated the generation of verbal commands for activating saved tasks. However, these tools entail usability challenges as they require users to manually record execution paths for each use case. In contrast, \tool innovates by predicting the most suitable action for each UI screen based on user requests without the need for pre-programming. Our database, tailored for streamlining task execution, is automatically constructed in real-time as users interact with their mobile apps.

\subsection{Large Language Models for Enhanced Human-AI Collaboration}
The advent of generative AI has given rise to innovative LLMs, such as GPT-4~\cite{openai2023gpt4}, DALL-E~\cite{ramesh2021zero} and Llama~\cite{touvron2023llama}. These LLMs have revolutionized the landscape of AI development by enabling developers to achieve complex tasks through few-shot prompting, eliminating the need for extensive custom model training. Their remarkable versatility has spurred active research in both IT and non-IT domains, spanning areas like software testing~\cite{liu2023fill, feng2023prompting}, high-performance computing~\cite{chen2023lm4hpc}, finance~\cite{wu2023bloomberggpt}, and health science~\cite{gilbert2023large}. LLMs have particularly excelled in enhancing the intuitiveness of existing methods, as seen in software testing, where they generate authentic text inputs based on the current UI page information, replacing the conventional random text input approach~\cite{liu2023fill}. This demonstrates the transformative potential of LLMs in advancing research and innovation across a multitude of domains.

The capabilities of LLMs have sparked a surge in their application within assistive technology, revolutionizing the translation of user commands into executable tasks across diverse systems. Recent research in this domain has witnessed the transformation of human natural language commands into various types of tasks, including visualization tasks~\cite{wang2023llm4vis}, operating system tasks~\cite{liu2023agentbench}, and robotic tasks~\cite{singh2023progprompt,liu2023lang2ltl}. LLMs have enabled these systems to tackle more intricate commands beyond the scope of existing heuristic approaches. They also exhibit a remarkable ability to comprehend variations of commands that share similar intentions but are expressed differently. This pioneering framework, with the support of LLMs, paves the way for a novel (semi)automated task execution paradigm, erasing the boundaries between traditional command patterns and intuitive command modalities.

\uist{Notable advancements include the World of Bits platform, which enables web-based agent training through interactions with real-world websites using low-level actions. WoB utilizes reinforcement learning and behavioral cloning to demonstrate these techniques' potential in web-based tasks, ensuring reproducibility through cached HTTP traffic~\cite{shi2017world}. Additionally, Mind2Web pushes the boundaries of generalist web agents by leveraging LLMs to handle complex, open-ended tasks across a wide range of real-world websites, showcasing the capacity of LLMs to generalize across diverse web environments~\cite{deng2024mind2web}. Generative agents further highlight the application of LLMs in simulating human behavior in interactive settings, providing a framework for more dynamic and lifelike simulations~\cite{generative2023agents}. Lastly, the design framework involving cells, generators, and lenses aims to optimize object-oriented interactions with LLMs, enhancing usability and functionality in various applications~\cite{cells2023framework}.}

Within the domain of mobile assistants, LLMs have become a transformative force, overtaking traditional machine learning models as demonstrated in previous works~\cite{voicify, autovci}. This shift has simplified the translation of natural voice commands into actions on mobile UIs. Wang et al.~\cite{wang2023enabling} used LLMs for conversation-like interactions with mobile UIs, showcasing a superior understanding of on-screen elements compared to earlier machine learning methods~\cite{li2020mapping}. However, their approach only focuses on single-screen support and interactions, which is inadequate for completing multi-step tasks. AutoDroid~\cite{wen2023empowering} and AutoTask~\cite{pan2023autotask} have employed LLMs, incorporating a degree of application knowledge (e.g., recalling previous actions or repeating similar commands) to execute multi-step tasks through a single command. that can complete multi-steps tasks under one command. Yet, these methods have tended to concentrate on discrete tasks without fully addressing the continuity between tasks within the same application. \tool advances this field by collecting sophisticated domain knowledge and employing advanced prompt techniques, aimed at improving precision and establishing a more advanced smartphone virtual assistant. This enhancement allows users to execute both familiar and novel tasks more effectively on their devices.

\section{The \tool System}

\begin{figure*}[]
\centering
\includegraphics[width=\textwidth]{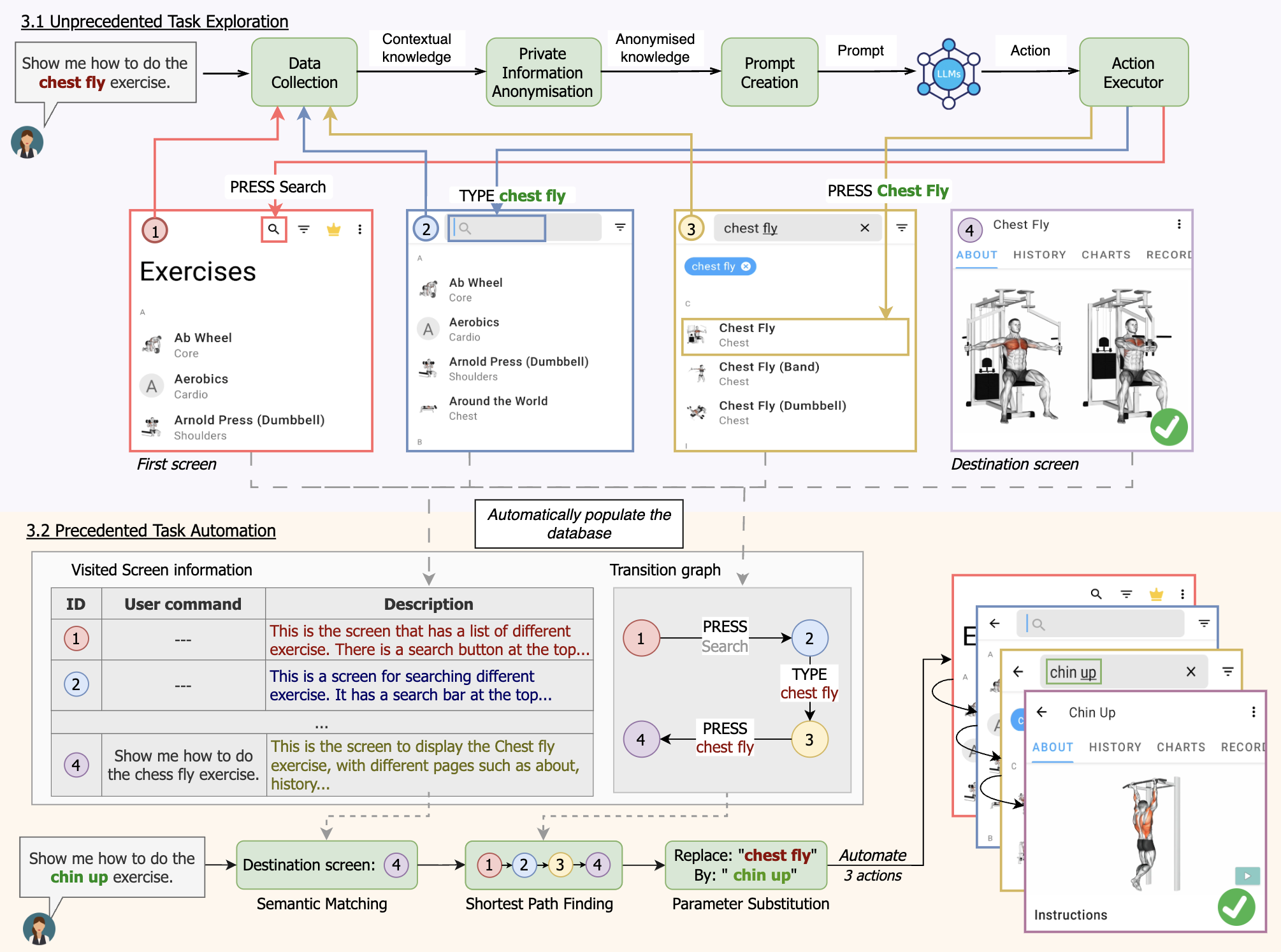}
\caption{An example use case in Home Workout application when the user needs to interact with the smartphone hands-free due to physical busyness. When performing an unprecedented tasks (Section~\ref{sec:onScreenInteraction}), \tool repeatedly predicts on-screen actions with current UI information and executes the response to achieve user tasks. The interactions collected during this process is then saved to streamline the execution of subsequent similar tasks (Section~\ref{sec:personalise}).
}
\Description{The figure shows 2 use cases in Home Workout application on smartphone, where the user is working out. The first use case shows on-screen example where the users says 'show me how to do the chest fly exercise', this command is passed in prompt engineering modules with current screen to form a prompt, which is then feed into Large Language Model. The response from Large Language Model is then passed to the Action Executor to perform a Press on search button, entering chest fly text and press on chest fly exercise in sequence to get to the destination screen. The second use case shows user executing the command ‘I want to do the chin up exercise, where this command is executed based on the saved tasks. First our tool find most semantically similar saved tasks, identify the shortest path from current screen to that destination screen using the transition graph. After that, our tool parameter substitution to fill ‘chin up’ into the exercise name and perform continuous execution to get to the chin up exercise.}
\label{fig:overview}
\end{figure*}

\label{sec:voicify}
We introduce \tool, a virtual assistant that empowers users to efficiently perform multi-step tasks on their smartphones using voice commands. Upon receiving a user command, \tool first attempts to streamline the task using the collected in-app navigation database (Section~\ref{sec:personalise}), to improve execution efficiency and reliability. If a task is unprecedented and not in our saved records, \tool will perform a series of step-by-step predictions of the on-screen navigation sequence to complete the task (Section~\ref{sec:onScreenInteraction}). Simultaneously, the system expands the database with new in-app navigation knowledge for subsequent autonomous task execution.

\subsection{Unprecedented Task Exploration}
\label{sec:onScreenInteraction}
Upon receiving an unprecedented task from the user, \tool will progressively predict and automatically execute each step until the task is accomplished. For each step, we collect contextual data from the mobile UI, task execution context, and current application information, which is combined with system-level information. \tool constructs all relevant data into prompts in a specific format and feeds them to the LLMs to determine the appropriate action on the user's smartphone. Upon receiving the response from the LLMs, \tool executes the action on the smartphone accordingly.

\subsubsection{Data Collection Module}
As LLMs execute logical reasoning based on textual inputs, known as prompts, a detailed and comprehensive prompt aids LLMs in understanding the task at hand and generating appropriate responses. Therefore, our primary focus is to incorporate sufficient information into our prompts to ensure accurate decisions from LLMs. This critical information, which we define as knowledge, is categorized into User Interface (UI) knowledge, Task knowledge, Application knowledge, and System knowledge. We extract this information through static analysis of the smartphone and its applications.

\textbf{UI knowledge.}
Information about the on-screen UI elements is a major component in our prompts, as it allows LLMs to comprehend the content currently displayed. Our primary emphasis is on representing smartphone GUIs in a textual format that can be interpreted by LLMs through text-based input. While recent research has proposed converting the UI elements list to HTML format~\cite{wang2023enabling,feng2023prompting} to reduce the prompt length, this approach becomes less relevant as LLMs now have relaxed restrictions on the number of tokens in a prompt. Therefore, we propose a more comprehensive view of the hierarchical structure of smartphone GUIs to improve the decision accuracy of LLMs. We represent each screen as a tree of nodes, with non-leaf nodes representing UI containers and leaf nodes representing visible UI elements. 
For each UI element, we collect the element type, text label, and append it with a unique ID. For some element types, such as buttons or text fields, the label can be extracted directly from the screen. However, for certain graphical UI elements like icons or image buttons, such information is not readily available. A potential solution is to apply deep learning models to predict the potential label of icons~\cite{chen2022towards}. Nevertheless, this approach can cause excessive overhead on app pages that include multiple images and icons, significantly impacting the responsiveness of real-time assistants. Alternatively, we propose a lightweight approach to collect alternative captions and resource names of these elements as labels, as they often include informative descriptions. For example, the search icon in screen 1 of Figure.~\ref{fig:overview} has the resource name \emph{``ic\_search''}, defined by app developers, indicating that this button is used for the search functionality. Furthermore, we collect the precise element location on the screen to cater to commands that refer to UI elements by their locations, such as \emph{``Press the icon at the top-right corner''}. We also retrieve the list of allowed actions for each element, which can include CLICKABLE, TEXT\_EDITABLE, SCROLLABLE, etc. This knowledge acts as a guardrail to ensure that LLMs do not return unsupported actions, such as pressing a disabled button.
The runtime UI elements may contain UI noise, which is a prevalent issue linked to the real-time gathering of UI elements~\cite{ui_noise}. This problem arises when the collected UI information does not align with its visual representation, which affects the semantic understanding of LLMs on current UI elements, resulting in incorrect interactions. To address this, we implement heuristics to mitigate potential inaccuracies and ensure the reliability of collected UI information. First, we utilise the collected coordination of each UI element to eliminate out-of-bound or empty elements. We also eliminate views are fully overlapped by other views, which does are invisible and not interactible. In addition, we remove those views that do not contain any interpretable information, such as empty view containers.


\textbf{Task knowledge.}
As complex tasks on smartphones involve multiple steps, treating each step as a separate action may lead to execution inaccuracies due to the misalignment of sequential actions. Additionally, the system may reattempt a single action multiple times, potentially causing an endless execution loop. To address this, we maintain information about the currently executing task and include it in our prompts. The task knowledge in each prompt specifies the user's request, the previously executed actions, and the visited pages. Such knowledge is incorporated into our prompts as natural language in a predefined template (e.g., \emph{``Started from <page 1>, we <action 1> <target 1> to get to <page 2>. After that, we <action 2> <target 2> to get to <page 3>.''}), helping LLMs to comprehend the previous actions and better align the subsequent actions accordingly.

\textbf{Application knowledge.}
Our experiment shows that LLMs can incorporate the knowledge about smartphone apps to output better results. This is particularly helpful for popular apps as LLMs are more familiar with these apps in their training data. In addition, these information helps validating each step in multi-step executions. For example, if the action caused unintended navigation to another app, LLMs can utilise the current app name to know this changes and output appropriate action to navigate back to the requested app. Therefore, we provide the app name, package name and list of activities as the application knowledge in our prompt. 
 
\textbf{System knowledge.}
To enhance the contextual understanding and reasoning capabilities of the LLMs, we not only fetch on-screen data but also retrieve relevant system-related information. We collect resolution information, which identifies the current screen orientation, and include it in our prompts. In addition, the screen resolution helps in the pre-processing of collected UI elements, such as validating the position of UI elements for out-of-bounds checking.

\subsubsection{Private Information Anonymisation.}
Smartphones often contain and display personal data, including phone numbers, addresses, and payment details. Anonymizing this data is crucial when inputting UI screen content into LLMs to prevent exposing sensitive user information. To this end, we have integrated a lightweight Named Entity Recognition (NER) to efficiently searches and categorizes textual information within the UI elements. Upon detecting private information, we substitute it with standardized tags, such as <address> or <phone number>, before the data is processed by the LLMs. This approach not only preserves user privacy but also ensures that LLMs understand the UI semantics accurately to generate appropriate responses.

\subsubsection{Prompt Creation}
\begin{figure*}[!h]
\centering
\includegraphics[width=\textwidth]{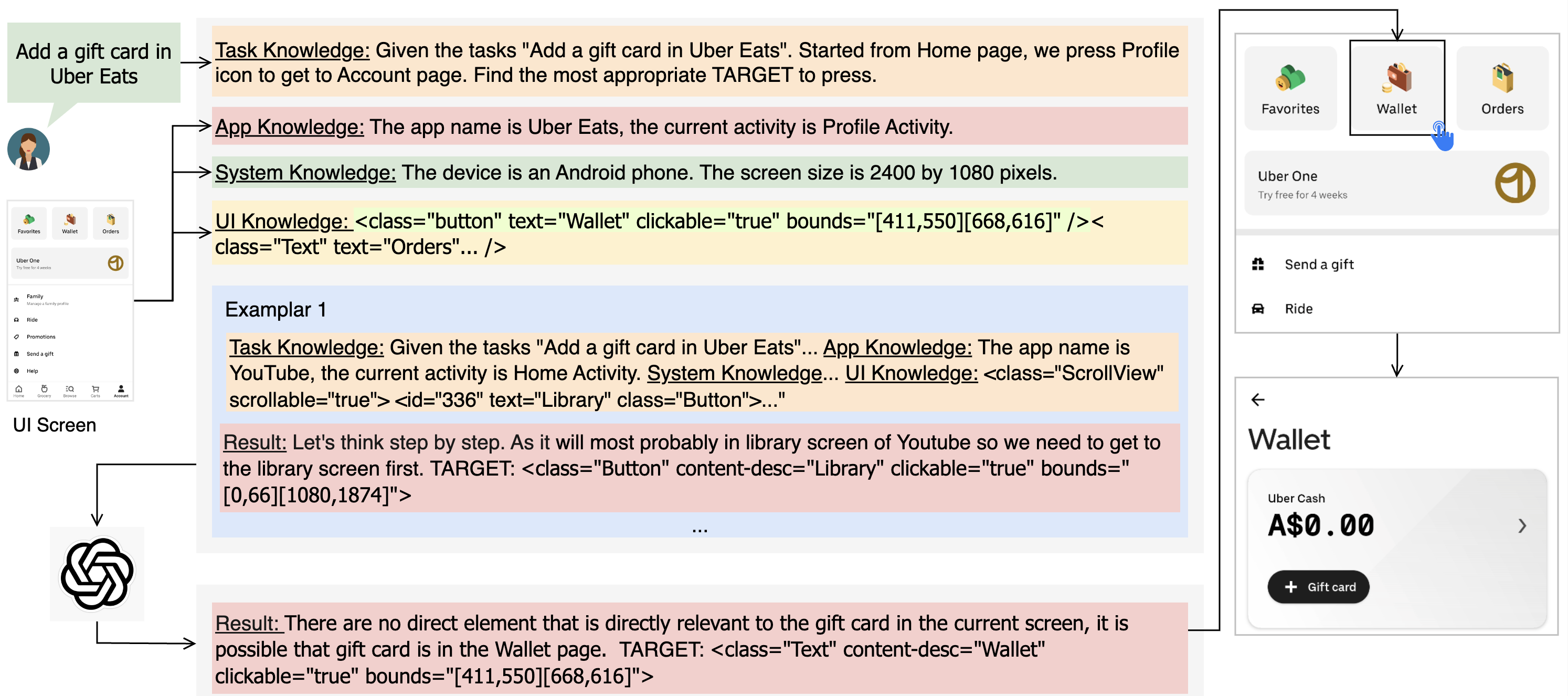}
\caption{An example of our prompt and response format to determine the most relevant target to press.}
\Description{In this figure, we show an example of our prompt to LLMs to determine the most relevant target to press. The prompt includes a task description, the user command, the application information, the UI elements on the current screen, and an exemplar that includes a prompt and response for LLMs to follow. For the response format, Large Language Model return the chain of thought and the result, which includes the information of the target such as the textual content and the bounds of UI element on the screen to perform a press.}
\label{fig:prompt}
\end{figure*}
We include the user command with collected data in the previous step in our prompt to predict the most suitable action to perform on user smartphone.  As the naive approach to prompt LLMs for various tasks may yield suboptimal results due to low accuracy and randomness in responses~\cite{chen2022program}, we adopt of different prompt engineering techniques. These techniques involve crafting prompts according to specific rules and components to elicit optimal responses from LLMs~\cite{liu2022design}.  Following Least-to-most Prompting strategy~\cite{zhou2022least}, \tool implements a two-step prompting approach to determine the action. First, we map the user task to a specific action (e.g., tapping an element, entering text, scrolling). Subsequently, based on the determined action, subsequent prompts are sent to identify the target UI element for executing that action ( as illustrated as an example Figure.~\ref{fig:prompt}). This approach allows breaking down a complex task into smaller steps, enabling the LLMs to improve accuracy.

To empower LLMs to facilitate rapid comprehension of specific rules and guidelines, we employed the Few-shot Prompting approach~\cite{brown2020language}, where we provide additional examplars in our prompt. Each examplar contains the sample prompt as the input with its corresponding response as the expected output. In addition, we integrate ``Chain of Thought''~\cite{wei2022chain} into our few-shot exemplars to help LLMs simulate human-like reasoning and provide logical output. This includes a sentence that explains the logic behind its output, which guides the LLM to apply a similar thought process in handling tasks. We dynamically calculate the number of examplars in a prompt based on the estimation of tokens used to include the knowledge. This approach maximizes the number of exemplars, thereby enhancing the accuracy of the response, while ensuring we do not exceed the token limit. In few-shot exemplar 1, as in Figure.~\ref{fig:prompt}, we provided a chain of thought that explains why Library button should be pressed to find the video history in YouTube. 

\subsubsection{Action Executor}
\label{sec:actionExecutor}
\tool extracts the action and target from LLMs response to perform the interactions on user's device, such as tapping, scrolling, or entering text on certain UI element. Prior mobile automation approaches~\cite{autovci} encounter challenges in handling runtime UI changes and app updates that might alter UI representations and operation sequences. To improve from previous approaches, \tool dynamically propose the action based on the current UI, ensuring reliable navigation through dynamic UI changes and updates. Additionally, \tool provides audio feedback to users, confirming that the system is automatically proceeding to the next command. This real-time feedback ensures a smooth and intuitive user experience with \tool's interaction capabilities. 

To address challenges in real-time system automation, including failure detection in automated steps~\cite{async_tool}, our implementation integrates a screen transition detector. This detector employs Hamming distance~\cite{hamming} to measure differences between screens, ensuring screen content changes due to the action. Moreover, to validate the success of actions, we implement additional heuristics. For example, in scenarios involving ENTER\_TEXT, we verify the presence of the entered text in the target text box. If an action proves inexecutable (i.e., not inducing appropriate changes to the UI), we repeat the step with supplementary information about the failed interaction attempts, thereby excluding these actions from the selection. Another challenge in mobile app automation is the dependency of UI screens on asynchronous internet content. Providing LLMs with incomplete or loading UI instead of fully rendered screen content may reduce the precision in identifying appropriate actions. In response, \tool delays UI collection until the screen is fully loaded. This is achieved by detecting screen-loading widgets in Android, leveraging information such as widget names, types, and shapes to identify progress bars and loading indicators. Additionally, we enhance our approach by collecting data on network transmissions to identify ongoing content download tasks, inspired by the methodology presented in~\cite{liu2020maddroid}, which leveraged network analysis for detecting ads. These methods enhance the reliability of \tool in effectively navigating through the app's interface.

\subsection{Precedented Task Automation}
\label{sec:personalise}

\begin{figure*}[!h]
\centering
\includegraphics[width=\textwidth]{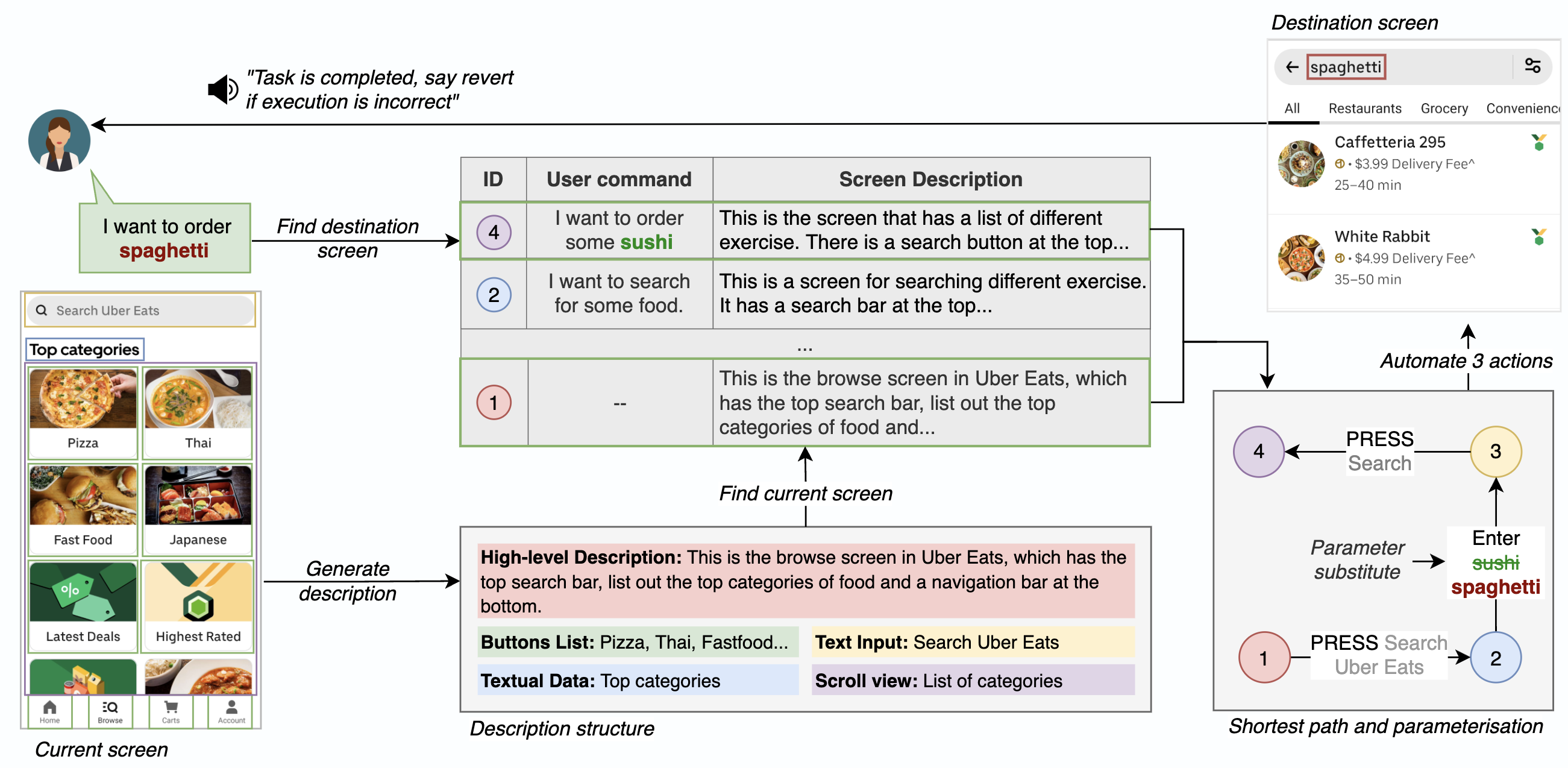}
\caption{An example use case in Uber Eats to how \tool use the historical tasks to execute user new command. The system first locate the current screen and destination screen from the collected graph. After that, it identifies and execute the action sequence to traverse to the destination screen. Finally, we utilise feedback from users to improve subsequent execution.}
\Description{An example use case in Uber Eats to how \tool use the historical tasks to execute user new command. Given an example task in Uber Eats, where user want to order some spaghetti. First, we search for similar command in the database and find the similar task that user ordered some sushi previously, which correspond to screen with ID is 4. Then we generate the description for the current UI elements on the screen, including a description, buttons list, text input textual data and scroll view. This description is used to find current screen in the graph, which is the screen with ID 1. After that, we find the shortest path from screen 1 to screen 4, including press search uber eats, enter spaghetti and press search. After the task being executed, the system give out audio output to confirms with users for task completion.}
\label{fig:database}
\end{figure*}

In mobile apps, UI elements on a specific app page and in-app navigation paths are predefined by developers during app development. Therefore, the series of user interactions on the screen to complete a task in one app remains consistent over different instances. Based on this, \tool automatically creates a saved path for each user request, enabling \tool to replicate interactions when receiving the same or similar commands from users. Unlike previous approaches~\cite{autovci,li2017sugilite}, which depend on manual task creation for automation support, \tool automatically records in-app navigation through on-screen interactions as described in Section~\ref{sec:onScreenInteraction}. This automated process not only expands the coverage of automated tasks but also eliminates the need for manual efforts in pre-defining shortcut tasks.  In this section, we outline our method (as shown in Figure.~\ref{fig:database}) to streamline subsequent similar tasks from users, combining both LLMs-based and heuristic-based modules. We introduce our database design in Section~\ref{sec:graph}. Initially, we identify the current UI screen displayed on the user's device and the destination screen (Section~\ref{sec:description}). We then find the most viable path from the current screen to the destination screen (Section~\ref{sec:executionGraph}). Finally, we incorporate human interactive feedback to validate and fine-tune the execution for future use (Section~\ref{sec:human_feedback}).

\subsubsection{Transition Graph}
\label{sec:graph}
In alignment with previous approaches for automating tasks in mobile apps~\cite{storydroid}, \tool utilizes a directed graph for each app to enable seamless transitions between different app pages. Each node in the database corresponds to a UI page in the app, containing a unique ID and a screen description, as detailed in Section~\ref{sec:description}. Additionally, for each node, we record the list of previous user requests that concluded on this page, as these indicate the functions served by the page.
The directed edges between nodes signify possible navigation paths from one UI page to another. These edges hold details about the required actions and target UI elements for page transitions. This data enables \tool to perform specific actions on the identified targets, thus replicating user actions to facilitate navigation between screens. This graph expands automatically to include new UI pages and associated page transitions as users interact with their smartphones.

\subsubsection{Screen Description \& Command Pattern Matching}
\label{sec:description}
Previous task automation approaches typically start from an application's launcher page~\cite{autovci,arsan2021app}, a method that falls short in real-world scenarios, especially when users request voice assistants to complete ongoing tasks. To address this, \tool enables task automation from any page of the app, aiming to fulfill any new or ongoing tasks from users. We first identify the node that represent the current app page in our graph database. However, identifying a page by performing string matching on the UI content presents substantial practical challenges. This difficulty arises as a specific application page can display dynamic content. For example, consider the search result page in the Uber Eats app, where distinct restaurants are displayed for \emph{``spaghetti''} and \emph{``sushi''} search terms. Despite featuring different UI content as they display different restaurants, these pages share the same layout and are identified as the same node in our graph database. To compare between app pages, \tool leverages the capabilities of LLMs to semantically summarize the UI content into a semi-structured description in natural language. The description includes a short paragraph describing overall functionality that the screen serves based on the UI elements, current activity name and app name. In addition, \tool appends the list of interactive elements, including clickable, scrollable, and text-editable elements to the description. To locate the current UI page among previously visited pages, \tool uses LLMs for semantic matching between the screen's description and existing screen descriptions in the database. As LLMs are proficient in logical tasks on natural language, the response allows us to identify the node in the graph database.

After identifying the current screen, we perform semantic matching to find the destination screen. We define the destination screen as the most relevant screen that can serve user requests. For example, the destination screen for user command \emph{``I want to order spaghetti''} is the search result page for the keyword \emph{``spaghetti''}, where users can view the list of spaghetti restaurants. We utilise the saved commands and screen description for each app page to prompt LLMs to determine the most relevant page from the app. We build a prompt that includes user requests and the list of saved screens in the app to rank the relevancy of each app screen with the request that user made. 
 
\subsubsection{Path Finding \& Execution.}
\label{sec:executionGraph}

After identifying the current and destination pages within the saved graph database, \tool uses the Shortest Path algorithm~\cite{golden1976shortest} to determine the sequence of actions required to navigate from the start to the destination node. \tool extracts this sequence from the edges connecting the start node to the destination node and executes each action in sequence using the Action Executor described in Section~\ref{sec:actionExecutor}.

Given the dynamic nature of smartphone GUIs, where the relative coordinates of buttons may vary due to scrollable screens or unexpected pop-ups and ads, execution validation is crucial. To ensure automation follows the expected sequence, \tool performs validation after each action. We generate the description for the current page (as explained in Section~\ref{sec:description}) and compare it with the description in the anticipated node. If the two descriptions do not semantically match, \tool reverts the preceding action and seeks an alternative path to the destination page. If no other path is found, \tool predicts and executes subsequent steps using the On-screen Interaction Module to explore a new path. This iterative process adapts to changes in the smartphone GUI, dynamically executing actions based on real-time screen information.

Additionally, \tool leverages command parameterization techniques~\cite{autovci, paramsmarcro}, allowing saved paths to be reused for similar tasks with different parameters. Using the LLMs' enriched vocabulary and robust natural language understanding, \tool prompts the LLMs to function as an advanced Named Entity Recognition system. This system identifies and replaces substitutable words with new keywords in the action sequence. For instance, in the Uber Eats app scenario (Figure.~\ref{fig:database}), a saved command for \emph{I want to order some sushi''} can be adapted by replacing \emph{sushi''} with \emph{spaghetti''}, thus modifying the command from ENTER \emph{sushi''} to ENTER \emph{``spaghetti''}. \tool then executes this adapted sequence to complete the task.

\subsubsection{Human Feedback Loop}
\label{sec:human_feedback}
One common issue with UI automation tools is dealing with changes in UI elements and in-app navigation due to new version updates by app developers. Additionally, pop-up ads may dynamically appear during the automation process. These challenges make predefined actions, such as clicking on an anticipated screen element becomes infeasible. To enhance the reliability of the saved paths, \tool incorporates a human-in-the-loop approach, modifying execution paths based on human feedback. After completing an autonomous task, \tool uses user feedback to validate whether the task was successfully executed. If users are unsatisfied with the automation, they can provide feedback in natural language, specifying where the issue occurs. This valuable information is saved and included in our prompts, resulting in improved accuracy for future executions. Additionally, we provide an interface for users to manually amend saved commands, allowing them to customize or shorten their commands to trigger certain automation according to their preferences. Through user feedback and iterative learning, \tool facilitates human-AI collaboration in the automation process, enhancing LLMs decision-making and personalizing user experiences.

\subsection{Implementation}
We developed \tool as an Android application, leveraging the Accessibility Service provided by the Android OS and programming it in Java \cite{accessibility_service}. Specifically, \tool is designed to subscribe to the \emph{typeWindowContentChanged} accessibility events \cite{accessibility_service}, enabling it to detect and respond to changes in the UI on the screen. We engineered a dynamic pipeline that systematically extracts UI elements and organizes them into a hierarchical structure. This is achieved by processing AccessibilityNodeInfo objects \cite{accessibilitynodeinfo}, which are Android's data representations of UI elements accessible through the Android Accessibility Service. Additionally, \tool gathers application-level information using the Android PackageManager class.

For the integration of a Large Language Model, we selected GPT-4, the most advanced model developed by OpenAI at the time of this research \cite{openAIAPI}. For personalized services, such as screen descriptions and transition graph data, we store this information locally on the device for efficient future access. We made \tool and the prompt templates\footnote{https://github.com/vuminhduc796/GPTVoiceTasker/blob/main/prompts.txt} publicly available at the GitHub repository\footnote{\url{https://github.com/vuminhduc796/GPTVoiceTasker}} for further research in this field.

\section{Technical Evaluation}
\label{sec:technicalevaluation}

To evaluate the effectiveness and reliability of the proposed system, we conducted three experiments on our command interpreting module, unprecedented tasks exploration and usage-based execution.
Specifically, we first assess the system's ability to comprehend user commands and perform on-screen interactions, comparing its performance to other state-of-the-art approaches. In addition, we experiment the ability of \tool to explore the execution path for unseen multi-step tasks. Lastly, we investigate the system's capability to execute multi-step tasks based on the saved user usages.

\subsection{On-screen Interaction Evaluation}
\label{sec:onScreenInteractionEvaluation}

\subsubsection{Experiment Setup \& Metric}

\textbf{Datasets:}
We collect a specialised test set to evaluate our system's capabilities in understanding natural language commands and mapping them to appropriate actions and target UI elements. This dataset comprises 278 natural language user commands to interact with on-screen Android UI elements.

Although prior research~\cite{voicify, burns2022motifvln} has produced a similar test set, it is not directly adaptable to our context for two critical reasons. First, some instances in the test set are artificially synthesized based on predetermined heuristic rules. The resulting natural language commands are linguistically biased toward simpler linguistic patterns and do not align with the complex linguistic variants inherent in real-world human spoken utterances. Second, some test examples in the existing dataset have become obsolete or are no longer replicable due to updates in the corresponding applications.

To construct a test set that more closely aligns with real-world user interactions, we adopted a data-driven approach. We engaged 31 participants (17 females, 14 males), with 4 individuals having never utilized voice assistants before, 4 using them 3-4 times a week, 6 using them daily, and 17 using them less than 3-4 times a week. All participants are work professionals and university students who use smartphones daily. We provided these participants with screenshots alongside a specific task to accomplish. We then recorded the verbal commands they issued to their mobile device to complete the given task. After the collection, we annotated the commands to specify the intended action and target UI elements within the Android system; here, the term \textit{action} refers to executable functions, while \textit{target} denotes specific UI components or elements on the current screen. As a result, we collected 278 natural user commands for the dataset.

\textbf{Metrics:}
Similar to Vu et al.~\cite{voicify}, we adopt three evaluation metrics, namely \textit{Exact Match Accuracy} (EM), \textit{Target F1} and \textit{Action F1}. 
EM calculates the percentage of instances in the test set where the predicted sequence exactly matches its corresponding ground-truth sequence.
The measures \(\textit{Target F1}\) and \(\textit{Action F1}\) quantify the average micro F1 score for the target (i.e., the UI components to be interacted with) and the action (i.e., the actions to be performed on the UI components), respectively. The F1 score for each instance is computed using the formula:
\[
\text{F1} = \frac{2 \times | \text{pred} \cap \text{gold} |}{| \text{pred} | + | \text{gold} |}
\]
\noindent where \(| \text{pred} |\) represents the size of the set of predicted targets or actions, and \(|\text{gold}|\) denotes the size of the set of ground-truth targets or actions. The average micro F1 score is calculated across all instances for either targets or actions.


\textbf{Baselines:}
\uist{We consider five baselines for converting natural language into semantic meaning representations, which comprise actions and targets. These baselines are \textbf{vanilla Seq2Seq}~\cite{bahdanau2015neural}, \textbf{BERT-LSTM}~\cite{xu2020autoqa}, \textbf{Voicify Parser}~\cite{voicify}, the \textbf{\tool w BasePrompt} model, and \textbf{Wang et al.}'s work~\cite{wang2023enabling}.} In the original work by Voicify, all three baselines employ deep learning models trained on datasets synthesized using the Overnight method~\cite{wang2015building}. This method generates training sets based on predefined lists of actions and targets that are designed for evaluation scenarios in Voicify. To ensure a fair comparison, we modified these lists to include the actions and targets present in our test dataset. We then re-synthesize the training set, which includes 1,384 instances, using the Overnight method, adhering to the implementation outlined in Voicify's work. 
\uist{The \tool with base prompt is used as an ablation study to the current prompt design of \tool. The base prompt contains only a minimal explanation of the task and the UI knowledge (XML file) of the current mobile screen.}
Wang et al.'s work~\cite{wang2023enabling} was the first to incorporate LLMs for interacting with mobile interfaces. We incorporated the 2-shots LLM prompts they demonstrated in the paper for mapping instruction to UI actions. Additionally, we have enhanced their model by integrating the more advanced GPT-4, which also inline with the one we used in \tool.

\begin{table}[]
\caption{The experiment results of different baselines compared with \tool in three metrics. }
\Description{
This table has 4 columns, first column represents the Model names as Sequence to sequence, BERT-LSTM, VoicifyParser and GPTVoiceTasker. The second to forth columns represent the exact match accuracy, Action F1 and Target F1. For Seq2seq, the exact match accuracy is 25.2\%, action F1 is 47.6\% and target F1 is 35.6\%. For BERT-LSTM, the exact match accuracy is 41.4\%, action F1 is 59.7\% and target F1 is 57.3\%. For VoicifyParser, the exact match accuracy is 47.5\%, action F1 is 64.0\% and target F1 is 58.8\%. For GPTVoiceTasker, the exact match accuracy is 84.7\%, action F1 is 91.7\% and target F1 is 84.7\%.}
\resizebox{0.5\textwidth}{!}{
\begin{tabular}{cccc}
\hline
\textbf{Models} & \textbf{EM Accuracy (\%)} &  \textbf{Action F1 (\%)} &  \textbf{Target F1 (\%)}\\ \toprule
\textit{Seq2Seq} & 25.2 & 47.6 &  35.6  \\ \hline
\textit{BERT-LSTM} & 41.4 & 59.7 & 57.3  \\ \hline
\textit{VoicifyParser} & 47.5 & 64.0  & 58.8  \\  \hline
\textit{\uist{\tool w BasePrompt}} & \uist{58.4} & \uist{71.6} & \uist{62.4}  \\  \hline
\textit{Wang et al.~\cite{wang2023enabling}} & 79.9 & 85.4  & 83.4 \\  \hline
\textit{\tool} & \textbf{84.7} & \textbf{91.7}  & \textbf{84.7} \\ \bottomrule
\end{tabular}}
\label{table:parser}
\end{table}

\subsubsection{Evaluation Result}
Table~\ref{table:parser} presents the results of our technical experiments. Overall, \tool outperforms all baseline models across all metrics, achieving an \textbf{84.7\%} EM accuracy, a \textbf{91.7\%} Action F1 score, and a \textbf{84.7\%} Target F1 score. Among the baselines without the LLMs, the Voicify Parser performs the best, aligning with the results reported in its original paper~\cite{voicify}. However, its performance suffers when faced with linguistic variations in our new test set. For instance, while the command ``\textit{back}'' is correctly interpreted as ``\textit{( PRESS , back )}'', the phrase ``\textit{return to last page}'', which represents the same command, is incorrectly parsed as ``\textit{( SWIPE , DOWN )}''. Both BERT-LSTM and Seq2Seq models encounter similar issues, largely because they share architectural and training similarities with the Voicify Parser, yet perform even worse due to Voicify Parser being specifically optimized for task completion on Android systems.
\uist{Despite the lack of prompt design, \tool with a base prompt achieved better performance than the other three DL-based baselines. However, when compared with \tool, the results indicate that incorporating a multi-level knowledge-based prompt design, along with few-shot learning and the Chain of Thought technique within \tool, can enhance the accuracy of converting natural language to on-screen actions and the corresponding elements.}
The method by Wang et al.~\cite{wang2023enabling} demonstrates the highest capability among the baselines, courtesy of the LLM's intervention, effectively rectifying the errors previously noted. However, the lack of the chain-of-thought and least-to-most prompt techniques occasionally leads to inaccuracies. This is evident in instances where the system misinterprets the intended direction in commands, such as confusing \emph{DOWN} with \emph{UP}, or when it cannot adequately differentiate between actions like \emph{PRESS} \emph{ENTER} or \emph{OPEN} when various verbs are employed in the commands.

Benefiting from the integration of LLMs and the prompting techniques, \tool excels at handling linguistic variants, consistently deriving the intended action and target regardless of variations in the input. The Action F1 score for \tool reaches 91.7, indicating its enhanced ability to predict actions across various linguistic patterns. Moreover, we observed that LLMs effectively learns the true associations between actions and targets, thereby excelling at target prediction as well. For instance, \emph{PRESS} is exclusively predicted with UI buttons, \emph{ENTER\_TEXT} is linked solely with text input fields, and \emph{OPEN} corresponds to app names. In contrast, the baselines often learns incorrect associations and outputs wrong target predictions. \uist{Our experimental results show that \tool is effective in understanding and accurately processing different linguistic variations, demonstrating its adaptability in real-world scenarios.}

\subsection{Multi-step Execution Evaluation}

\subsubsection{Experimental Setup \& Metrics}
This experiment assesses the Unprecedented Task Exploration module's capability to execute unseen tasks. We evaluated the module's performance using the most recent human-collected demonstrations and natural language instructions from the Android-in-the-wild dataset~\cite{rawles2024androidinthewild}. This dataset provides step-by-step on-screen interactions to complete tasks based on natural language instructions, mirroring real-world commands. From this dataset, we randomly sample the data from multi-step Google Apps subset and manually validate each pair to get 140 test cases, with each test case has from 1 to 15 steps (M=6.705, SD=2.764). We treat the end screen after the final action in the action sequence from the dataset as the ground truth, indicating successful command execution. We did not evaluate the accuracy of each steps to the dataset as one task could be performed by multiple approach. For each test case, we have \tool iteratively explore the path to complete the instruction, comparing the destination screen achieved by \tool with the dataset's ground truth. A test case was deemed successful if \tool reached the same screen as it is in the ground truth with no more than three additional steps than the dataset demonstration. Cases where the step count was exceeded or the next step could not be identified were marked unsuccessful. \tool executed commands solely using its task exploration module, without relying on a database for guidance. Similar to Sec~\ref{sec:onScreenInteractionEvaluation}, we used Wang et al.~\cite{wang2023enabling} approach as the baseline, which used a 2-shot prompting technique on the GPT-4 model, making iterative requests after each action response. 

\subsubsection{Results}
 \tool achieved an \textbf{85.7\%} success rate (120 out of 140), outperforming the baseline approach by Wang et al., which achieved a \textbf{56.4\%} success rate (79 out of 140). \tool succeeded in performing logical reasoning to execute the tasks. It utilizes the current app name and package name to determine if required apps need to be opened, and uses the list of run-time device-available app names to query the most suitable app for the task. Notably, \tool effectively navigated tasks without encountering cyclic navigation issues or repeating actions, such as repeatedly tapping the \emph{Settings} title within the Settings app, which hindered task completion in the baseline. This improvement is attributed to the integration of historical messages and runtime execution error handling module. Our study also found that \tool does not rigidly adhere to the demonstrated paths presented in the Android-in-the-wild dataset for task completion. For example, in managing tasks within the Android Settings App, it occasionally opted to search directly for options rather than scrolling through menus to locate them. Other added constraints in \tool helped to improve the usability of multi-step interactions, such as validating if the current screen is scrollable or a text field is focused before inserting text, which were observed as reasons that caused failures in the baseline.

Further investigation into unsuccessful cases highlighted two main areas for improvement. First, \tool struggled with time-related tasks, such as \emph{``Check the schedule for Friday next week''}, due to the LLM's limited knowledge of the current date and time. As a result, while \tool correctly opened the app, it was unable to select the Friday of next week to view the schedule. This could be mitigated by integrating temporal information into our prompts. Second, \tool exceeded the number of steps when performing \emph{``Open the settings page in Google Maps''}, where the immediate step required pressing the profile picture, which is not directly relevant to the task command. To address this issue, we could enrich the prompts with additional contextual knowledge, such as the steps involved in accessing settings in similar applications where pressing the profile picture is necessary. Additionally, incorporating a strategy of random exploration steps before repeating a command could help the system discover more direct paths to complete tasks.

\subsection{Database Execution Evaluation}

\begin{table}[htbp]
    \centering
    \caption{Saved task execution evaluation result for direct match tasks and parameterised tasks across 5 categories.}
    \Description{
    This table shows the average number of automated steps, success rate for direct match and success rate for parameterised tasks for each app categories. For message friends app, the average number of steps is 4.33, achieved 93.33\% for direct match accuracy and 86.67\% accuracy for parameterised task. For listen to music app, the average number of steps is 5.27, achieved 80\% for direct match accuracy and 73.33\% accuracy for parameterised task. For set an alarm app, the average number of steps is 5.73, achieved 73.33\% for direct match accuracy and 53.33\% accuracy for parameterised task. For check weather app, the average number of steps is 5.07, achieved 80\% for direct match accuracy and 73.33\% accuracy for parameterised task. For get direction and maps, the average number of steps is 5.53, achieved 86.67\% for direct match accuracy and 73.33\% accuracy for parameterised task. On average, the average number of steps is 5.19, achieved 82.67\% for direct match accuracy and 72\% accuracy for parameterised task.
    }
    \label{tableautomated}
    \resizebox{0.5\textwidth}{!}{
    \begin{tabular}{c|c|c|c}
        \hline
        \textbf{Category} & \begin{tabular}[c]{@{}l@{}}\textbf{Average Number of} \\ \textbf{Automated Steps}\end{tabular} & \multicolumn{2}{c}{\textbf{Success Rate (\%)}} \\
        \cline{3-4}
        && \textbf{Direct Match} & \textbf{Parameterised} \\
        \hline
        Message Friends & 4.33 & 93.33 & 86.67 \\
        \hline
        Listen to Music & 5.27 & 80.00 & 73.33 \\
        \hline
        Set an Alarm & 5.73 & 73.33 & 53.33 \\
        \hline
        Check Weather & 5.07 & 80.00 & 73.33 \\
        \hline
        Get Directions \& Map & 5.53 & 86.67 & 73.33 \\
        \hline
        \textbf{Average} & \textbf{5.19} & \textbf{82.67} & \textbf{72.00} \\
        \hline
    \end{tabular}}
\end{table}

\subsubsection{Experimental Setup \& Metric}
    
 In this experiment, we assessed \tool's ability to automate tasks using the usage-based execution module. We initially identified the five common smartphone application categories, as shown in previous study~\cite{arsan2021app}. Within each application category, we randomly selected five popular applications from the Google Play Store, with downloads ranging from 1 million to over 1 billion. For each selected app, we identified three features introduced by the developers in their Play Store descriptions. Each feature was then used to create both a direct match task, involving a straightforward match between user commands and corresponding app actions, and a parameterized task, requiring \tool to perform keyword substitutions to complete the task successfully, as shown in Section~\ref{sec:executionGraph}. For creating the direct match test cases, we paraphrased each saved command using state-of-the-art paraphrasing tool Quillbot\footnote{\url{https://quillbot.com/}}, as in~\cite{shiri2022paraphrasing}. In the case of parameterized tasks, we substituted one entity in the paraphrased command with another entity that has similar semantic. For example, consider the saved task ``\emph{Get directions to the nearest supermarket}''. In this case, the direct matching task would be ``\emph{Find the nearest supermarket's location}'', while the parameterized task would involve substituting ``\emph{restaurant}'' for ``\emph{supermarket}'', resulting in ``\emph{Find the nearest restaurant's location}''. This process resulted in a total of five app categories, each category contains 15 direct match tasks and 15 parameterised tasks. These tasks involve 4 to 7 steps, with an average of 5.19 steps per task as illustrated in Table~\ref{tableautomated}. All tasks can be automated with one voice command with the saved user app usage patterns. For a detailed list of the apps and features used in the experiment, please refer to our GitHub repository\footnote{\url{https://github.com/vuminhduc796/GPTVoiceTasker/blob/main/Result.xlsx}}.

To populate the transition graph and store screen descriptions, we manually navigated through each screen in every application using \tool. Subsequently, we configured the saved commands to reach the respective screens as the ground truth. We used the success rate as the primary metric, each test case is marked as success if \tool can successfully opened the desired feature using a single command.

    \subsubsection{Results}
Table~\ref{tableautomated} illustrates the accuracy of our saved task execution modules. Our findings indicate that \tool achieved an impressive level of automation, successfully handling \textbf{82.7\%} of exact match tasks and \textbf{72.0\%} of parameterized tasks. Notably, \tool exhibited exceptional performance in tasks related to messaging and directions \& maps applications. This success can be attributed to the relatively static nature of these apps, where user interfaces maintain a consistent structure. Our results underscore \tool's proficiency in command analysis, semantic matching to saved tasks, and parameterized phrase substitution within these contexts. However, the accuracy of \tool diminished when confronted with tasks related to setting alarms. To better understand the root causes of this decline in performance, we conducted an error analysis on the failed test cases. Several key issues emerged:

\begin{itemize}
    \item \textit{Complex Parameterized Tasks}: For parameterized tasks with additional steps, such as setting an alarm for 7:30 instead of 7:00, \tool struggled due to the extra step involved in selecting the minutes, which was on a separate UI element. Further works include making \tool adaptable to these additional steps in the automation process.
    
    \item \textit{Pop-ups Ads and Unusual UI Elements}: Certain applications presented pop-ups ads and unusual UI elements in run time that were not encountered during the initial task-saving process. Consequently, \tool faced difficulties in completing these tasks. To improve the robustness of our approach, we recommend exploring the integration of a deep learning model to detect and handle such ad widgets and unusual UI elements, as in~\cite{liu2020maddroid, dong2018frauddroid}.
\end{itemize}

\section{User Study}
\label{sec:evaluation}
To demonstrate the practical utility of our tool, we conducted a user study to evaluate the holistic performance of the \tool system within real-world scenarios. Our evaluation involved a comparative analysis against two baseline systems: 1) Voice Access~\cite{yamada_2020}, the official voice assistant product developed by Google, with over 100 million downloads, and 2) Voicify~\cite{voicify}, the state-of-the-art research product endeavor incorporating deep learning models to enhance command comprehension. This study pursued a threefold objective: 
i) establish a performance benchmark for user interactions utilizing the \tool system as opposed to the aforementioned baseline systems, 
ii) juxtapose user feedback concerning the cognitive load and overall usability of the \tool system against the baselines 
and iii) capture qualitative insights from participants, thus enabling the identification of potential avenues for enhancing the \tool system. In order to achieve these objectives, we recorded the task completion times for tasks undertaken using both the \tool system and the baselines. Furthermore, a comprehensive post-experiment interview was conducted with each participant, facilitating the collection and analysis of both quantitative and qualitative feedback.

\begin{table*}[h]
\caption{The list of tasks for user evaluation.}
\Description{This table displays the list of tasks for user evaluation. Task 1 is ‘Check, send, and delete messages’ in messages app, with has 8 steps. Task 2 is ‘Search for a specific song and play it’ in Apple Music, with has 8 steps. Task 3 is ‘Create a new alarm and save it’ in Challenges Alarm Clock app, with has 10 steps. Task 4 is ‘Check the weather within a particular city’ in BOM Weather app, with has 6 steps. Task 5 is ‘Write and delete a note’ in Notes app, with has 8 steps. Task 6 is ‘Search for a pizza store, and complete the order’ in Uber Eats app, with has 10 steps.}
\centering
\begin{tabular}{c|l|c|l|c}
\hline
\textbf{No.} & \multicolumn{1}{|c|}{\textbf{Task}} & \textbf{\#Steps} & \multicolumn{1}{|c|}{\textbf{App Name}} & \textbf{\#Downloads} \\ \hline

1 & Check the weather within a particular city. & 6 & BOM Weather & 1M+ \\ \hline
2 & Search for a specific song and play it. & 6 & Apple Music & 100M+ \\ \hline
3 & Create a note and write "Hello world" and delete it. & 8 & Notes & 1M+ \\ \hline
4 & \parbox[t]{6cm}{Check for an unread message, reply with a message and delete the conversation.} & 8 & Messages & 1B+ \\ \hline
5 & Search for a pizza store, and complete the order. & 10 & Uber Eats & 100M+ \\ \hline
6 & Create a new alarm and save it. & 10 & Challenges Alarm Clock & 1M+ \\ \hline

\end{tabular}
\label{table:tasks}
\end{table*}

\subsection{Tasks}

We designed 6 experimental tasks, encompassing a broad spectrum of the most common interactions performed on the screen, ranging from tapping and swiping to entering text. Each task was structured to comprise between 6 to 10 sequential steps. The detailed list of these tasks is outlined in Table~\ref{table:tasks}.

\subsection{Participants}

We recruited 18 participants, consisting of 10 males and 8 females, aged between 18 and 31 years old for our study. \uist{The mean age of participants was 25.83 years (SD = 4.26). The group included 8 bachelor students (from IT and Business fields), 4 master students (IT), and 6 PhD candidates.} 8 participants are native English speaker while all other participants are proficient in English. All participants possess a commendable level of familiarity with technological devices and actively use smartphones in their daily routines. 

\uist{We advertised our experiment on LinkedIn to recruit participants from our university. During recruitment, participants provided their gender, age, study level, English proficiency, tech proficiency, and experience with voice assistants.}
While participants exhibited exposure to virtual assistants like Siri or Google Assistant, none were acquainted with utilizing assistive tools for smartphone control via voice commands. Specifically, none of the participants had prior experience with any of the experimental tools employed in our study. This participant selection was deliberate, as our study sought to gauge the learnability aspect of the experimental tools. Each participant received a USD \$30 gift card for the participation.

\subsection{Procedure}
We conducted face-to-face user evaluations using an Android device as the experimental tool. On this device, we had the graph of each experimental app populated, which include the majority of app pages and navigation within the app. At the start of the sessions, participants were introduced to all experimental tools via demonstrative videos. The preliminary phase involved practicing basic tasks across all tools, enhancing participants' familiarity with step-by-step instructions and informative walk-through videos. We also use the searching for exercise tasks in Figure.~\ref{fig:overview} as the practice tasks, allowing users to achieve this task using each of the tool.

After that, participants independently executed six distinct tasks with no experimenter intervention. Each tool was employed for the completion of two tasks, and participants remained unaware of which tool was developed by us. To mitigate any potential biases, the order of tasks and the tools used were systematically counterbalanced for each participant~\cite{depuy2014counterbalancing}.

We applied a time cap of 60 seconds per step. We recorded the time taken to fulfil each task, including the cut-off time to perform quantitative analysis. We collected 108 data entries since each of the 18 participants has finished 6 tasks. In the end, using the System Usability Scale (SUS)~\cite{bangor2008empirical} form with a 5-point Likert scale, we evaluate the usability of \tool, compared to Voice Access and Voicify.
In addition, we investigated the cognitive load when experimenting with each tool using the NASA-TLX~\cite{hart2006nasa} form with a 7-point Likert scale. Lastly, we collected qualitative feedback on which part they liked the most about \tool and what might improve the system.

\subsection{Result}
\label{sec:result}
    \begin{figure}
    \centering
    \includegraphics[width=\linewidth]{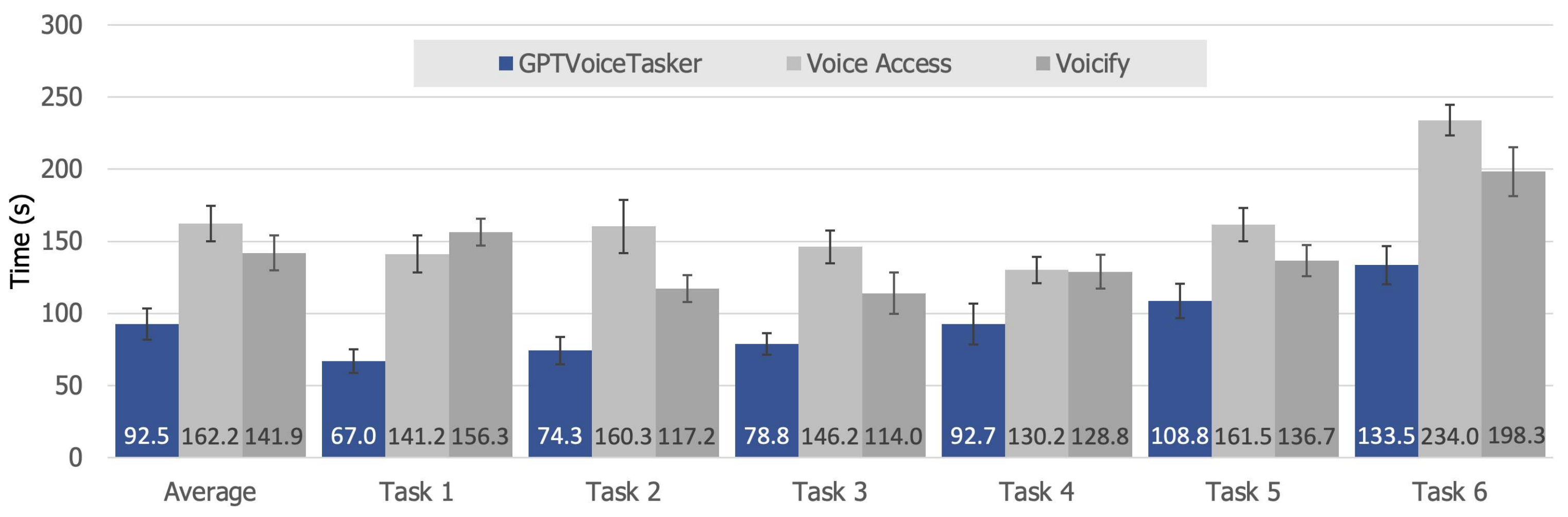}
    \caption{The average time taken to complete each task using \tool and the baselines in seconds.}
    \Description{
    The figure shows the average time taken to complete each task of GPTVoiceTasker, Voice Access and Voicify. The average time to complete the all tasks with GPTVoiceTasker is 92.5 seconds, while the 2 baselines achieved 162.2 seconds and 141.9 seconds. For task 1, the average time taken to complete the task for GPTVoiceTasker is 78.8 seconds, while Voice Access is 162.2 second and Voicify is 141.9 seconds. For task 2, the average time taken to complete the task for GPTVoiceTasker is 92.7 seconds, while Voice Access is 146.2 second and Voicify is 114 seconds. For task 3, the average time taken to complete the task for GPTVoiceTasker is 108.8 seconds, while Voice Access is 161.5 second and Voicify is 136.7 seconds. For task 4, the average time taken to complete the task for GPTVoiceTasker is 67 seconds, while Voice Access is 141.2 second and Voicify is 156.3 seconds. For task 5, the average time taken to complete the task for GPTVoiceTasker is 160.3 seconds, while Voice Access is – second and Voicify is 117.2 seconds. For task 6, the average time taken to complete the task for GPTVoiceTasker is 135.5 seconds, while Voice Access is 234 second and Voicify is 198.3 seconds.
    }
    \label{fig:data}
    \end{figure}

    \begin{table*}
    \centering
    \caption{Average number of automated steps by all participants in each task.}
    \begin{tabular}{c|c|c|c|c|c|c|c}
     
        \hline
        & \textbf{Average} & \textbf{Task 1} & \textbf{Task 2} & \textbf{Task 3} & \textbf{Task 4} & \textbf{Task 5} & \textbf{Task 6} \\
        \hline
    
        \textbf{\#Steps Automated} & 2.22 & 2.67 & 2.00 & 1.67 & 2.17 & 2.50 & 2.33 \\
        \hline
    \end{tabular}
    \label{tab:stepSkipped}
\end{table*}

\subsubsection{Overall User Performance}    
\begin{figure*}[!h]
    \centering
    \includegraphics[width=0.9\linewidth]{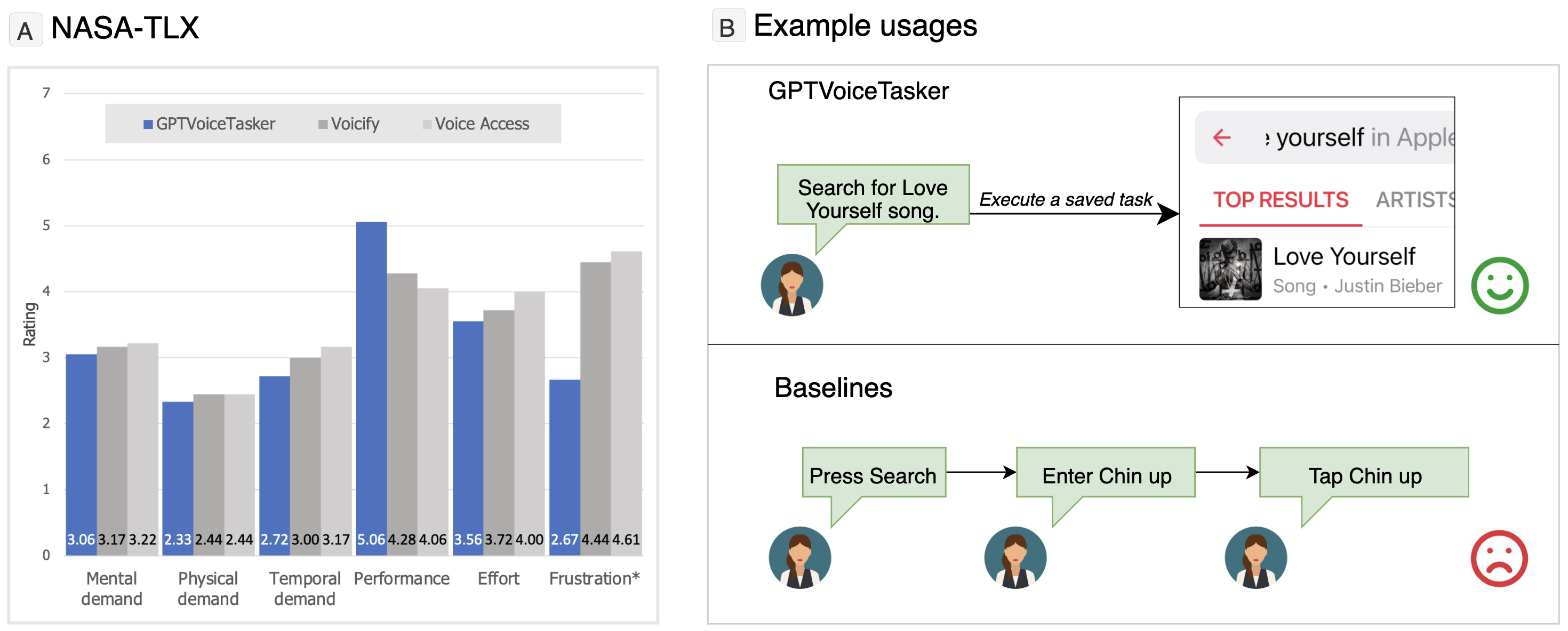}
    \caption{The comparison between \tool, Voicify, and Voice Access for A) the average cognitive load when using NASA-TLX form (lower is better) *: p < 0.01, **: p < 0.001 and B) Task 2 from the user evaluation with \tool and other baselines.}
    \Description{
    This figure contains 2 parts A and B. For part A, the chart displays the NASA-TLX result for GPTVoiceTasker, Voicify and Voice Access on the scale from 1 to 7, with the lower means better for all values, except for the performance. The mental demand for GPTVoiceTasker is 3.06, while Voicify achieved 3.17 and Voice Access achieved 3.22. The physical demand for GPTVoiceTasker is 2.33, while Voicify achieved 2.44 and Voice Access achieved 2.44. The temporal demand for GPTVoiceTasker is 2.72, while Voicify achieved 3 and Voice Access achieved 3.17. The performance for GPTVoiceTasker is 5.06, while Voicify achieved 4.28 and Voice Access achieved 4.06. The effort for GPTVoiceTasker is 3.56, while Voicify achieved 3.72 and Voice Access achieved 4. The frustration for GPTVoiceTasker is 2.67, while Voicify achieved 4.44 and Voice Access achieved 4.61. Part B include an example of a task to search for love yourself song using GPTVoiceTasker to directly execute the task, while it requires the baselines 3 steps to complete.
    }
    \label{fig:chart}
\end{figure*}

In Figure.~\ref{fig:data}, we present the average task completion times for each experimental tool. Our tool stands out with an average completion time of \textbf{92.5} seconds, surpassing Voice Access (162.2 seconds) and Voicify (141.9 seconds). This improvement in \tool's performance can be attributed to two primary factors. 
Firstly, we can tell that \tool is better at comprehending user intentions and mapping user commands to the correct actions on specific UI elements, regardless of the command format. In contrast, baseline tools often demand specific command formats, introducing errors in various usages. This issue caused extra time costs as participants needed to seek different ways to express their intentions with the baseline tools. For example, participants tried to tap the option button, in the Notes app with Voice Access by multiple attempts such as ``\emph{press on the option button}'', ``\emph{press the three-dot icons}'', ``\emph{tap icon for options}'' before successfully give the right command ``\emph{tap option}''.
Secondly, \tool optimizes the performance by automating several steps in one user command, as shown in Table~\ref{tab:stepSkipped}. On average, the participants saved 2.2 steps across all six tasks.
For instance, in Task 2, \tool efficiently automated the process of searching for Love Yourself song (as in Figure.~\ref{fig:chart}(B)), drawing from a previously stored action designed for searching other songs. 
This eliminated the need for three steps required for in-app navigation. However, some participants did not realize that they could trigger the saved tasks, leading to a missed opportunity for a significant performance boost.
In addition, \tool relates to network latency when sending and receiving data from the LLMs API endpoint. This issue could be mitigated with a better network connection.
    
\begin{figure*}[!h]
    \centering
    \includegraphics[width=0.95\linewidth]{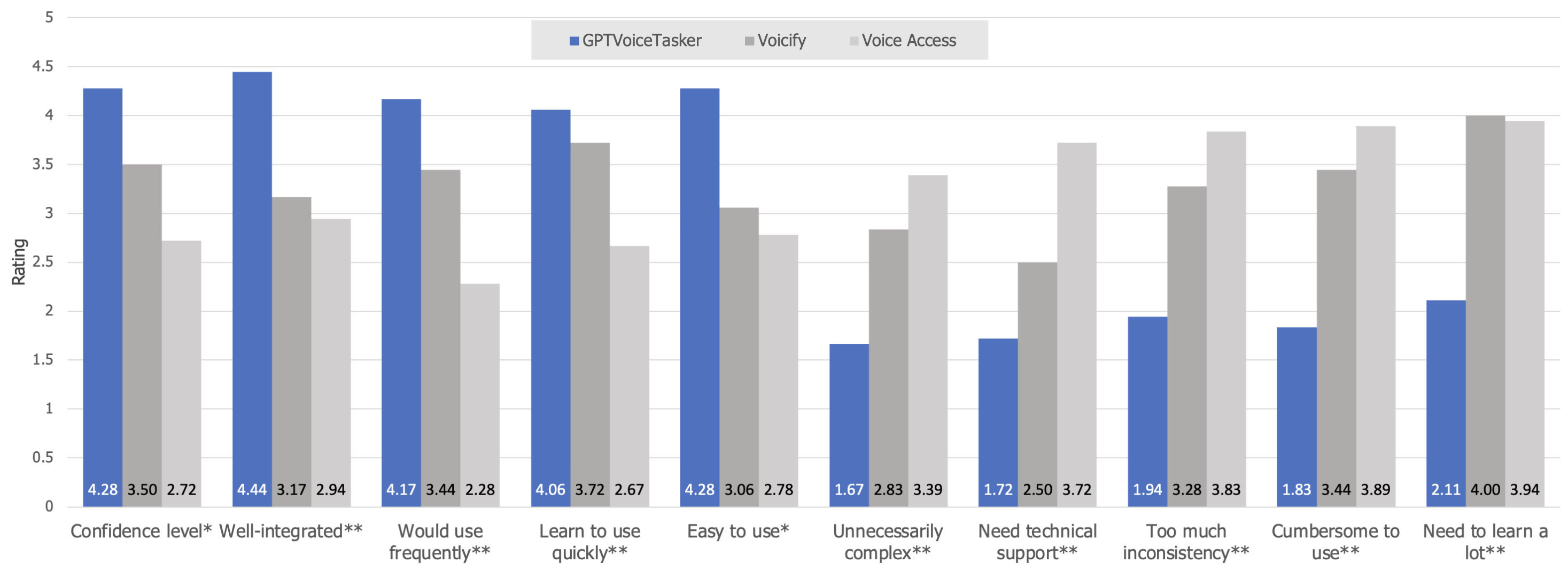}
    \caption{The comparison between \tool, Voicify, and Voice Access for the System Usability Scale (SUS). *: p < 0.01, **: p < 0.001.}
    \Description{The figure shows the System usability scale result for GPTVoiceTasker and 2 baselines on the scale 1 to 5, where higher is the better for the first five category while lower is better for the last 5 categories. For confidence level score, GPTVoiceTasker achieved 4.28, while Voicify score is 3.5 and Voice Access score is 2.72. For well-integrated score, GPTVoiceTasker achieved 4.44, while Voicify score is 3.17 and Voice Access score is 2.94. For would use frequently score, GPTVoiceTasker achieved 4.17, while Voicify score is 3.44 and Voice Access score is 2.28. For learn to use quickly score, GPTVoiceTasker achieved 4.06, while Voicify score is 3.72 and Voice Access score is 2.67. For easy-to-use score, GPTVoiceTasker achieved 4.28, while Voicify score is 3.06 and Voice Access score is 2.78. For unnecessarily complex score, GPTVoiceTasker achieved 1.67, while Voicify score is 2.83 and Voice Access score is 3.39. For need technical support score, GPTVoiceTasker achieved 1.72, while Voicify score is 2.5 and Voice Access score is 3.72. For inconsistency score, GPTVoiceTasker achieved 1.94, while Voicify score is 3.28 and Voice Access score is 3.83. For cumbersome score, GPTVoiceTasker achieved 1.83, while Voicify score is 3.44 and Voice Access score is 3.89. For need to learn a lot score, GPTVoiceTasker achieved 2.11, while Voicify score is 4 and Voice Access score is 3.94.}
    \label{fig:chart2}
\end{figure*}
        
\subsubsection{Cognitive Load \& Usability Ratings}
Figure.~\ref{fig:chart}(A) presents an overview of participant feedback regarding their cognitive load levels for each system, assessed using the NASA-TLX form. \uist{We conducted a Friedman test~\cite{friedmantest} for statistical analysis on the result}. Participants reported decreased mental demand, temporal demand, and effort when using \tool in comparison to the baseline systems while achieving better performance. This result show an improvement in \tool's ability to reduce the cognitive load required for operation, aligning with our design goal.
    
To assess \tool's usability in comparison to the baseline systems, we employed \uist{Friedman test for statistical analysis} on collected System Usability Scale (SUS) scores, as depicted in Figure.~\ref{fig:chart2}. The analysis verified the enhanced usability of the voice control system, with \tool achieving an average SUS score of 79.861, surpassing Voicify (47.917) and Voice Access (36.528). \uist{Participants found \tool less complex (p < 0.001), less inconsistent (p < 0.001) and well-integrated (p < 0.001), leading to more frequently use (p < 0.001). In addition, participants can learn to use \tool quickly (p < 0.001) as they do not need to learn a lot (p < 0.001).} This remarkable outcome can be attributed to \tool's ability to effortlessly comprehend natural human commands, reducing the need for extensive training and practice. The lower likelihood of misinterpreting user commands also contributed to the positive results.
    
\subsubsection{Qualitative Feedback}
In this section, we collate qualitative feedback from participants after the experiment. Overall, the participants are satisfied with the tool, as well as providing suggestions for further improvements.
     
\textit{Ability to precisely interpret and execute human command.}
Participants expressed enthusiasm about the remarkable ability of \tool to interpret human commands naturally, enhancing the overall system's intuitiveness. P1 and P12 highlighted that they could issue commands \emph{``in their preferred manner''} and \emph{``converse naturally''} with \tool. This addresses cognitive overload concerns, as P4 appreciated the \emph{``stress-free experience''}, and P6 and P7 found \tool more \emph{``comfortable to use''}. For instance, when adding a new note, users could simply say \emph{``add a new note''} to prompt \tool to press the add button on the screen. 
Moreover, participants were impressed by our tool's accuracy in handling user input errors. P3 noted their satisfaction with how \tool \emph{``can still execute the correct action even when I make mistakes in my commands''}. Both P4 and P17 highlighted the tool's usefulness in daily tasks, as it eliminates the need to \emph{``exercise caution and stay alert''} when interacting with \tool. These feedback remarks strongly affirm the practicality of our approach in real-world task scenarios. In contrast,  traditional approaches typically demand fixed input formats, making them ill-suited for real-world scenarios where user input can vary significantly.

\textit{Automated execution helps accelerate tasks and improve user experiences.}
Participants offered positive feedback regarding the use of saved task automation, highlighting its significant impact on efficiency and user experiences. P11 mentioned that this feature is \emph{``accelerating the tasks''} while P13 emphasized the potential utility of \tool during physical activities, stating it would be \emph{``really useful when I work out''}. P5 appreciated this feature, describing it as \emph{``perfect for voice-interacting tools''}, as it mitigates the inherent challenges of voice command interactions. Additionally, P18 praised the feature, noting that tasks became \emph{``fairly easy''} with its implementation, indicating significant performance improvements. This, combined with the advanced capability to understand user intentions, enhances the intuitiveness of voice-based interfaces. When using a smartphone, users often have a specific task in mind, such as setting an alarm or checking the news. Unlike other approaches that require users to perform additional steps to translate their intention into executable commands that a voice interface can understand and execute, \tool can directly execute these tasks without causing additional mental stress.
However, users also provided valuable suggestions for enhancement. They expressed the desire for \tool to suggest executable saved tasks and display a list of saved tasks. Furthermore, participants suggested improving the introduction of this feature, as P4 noted it was \emph{``not familiar at first''}, and P6 emphasized the need for \emph{``better introduction.''} These insights underscore opportunities to refine the feature's usability and user onboarding, ultimately enhancing overall user satisfaction.
    
\textit{Suggestions for enhancing user experience.} 
Participants provided valuable suggestions for improving the intuitiveness of \tool. Regarding UI design, P14 recommended the inclusion of a \emph{``live transcription''} feature to display recognized voice commands. This would help users confirm that their commands were correctly received and make necessary adjustments if needed. Furthermore, P1 and P15 suggested incorporating a \emph{``loading indicator''}  to signify ongoing executions, addressing latency issues caused by execution delays. In terms of functionality, P7 proposed displaying a list of available tasks as suggestions, enhancing user interaction. Additionally, P15 discussed the potential for an interface that allows users to modify saved tasks, providing greater customization. Lastly, participants P7 and P12 suggested making the audio feedback from \tool clearer. These suggestions hold significant value for \tool's continuous improvement, aiming to deliver a more seamless user experience.

\section{Discussion}
\label{sec:discussion}

We introduced \tool as an autonomous speech-based virtual assistants. In this section, we delve into the implications and limitations of \tool.

\textit{Towards the adoption of the voice-centric interface.}
The advancements in natural language understanding, particularly through LLMs like GPT and Bard~\cite{chatgptandbard}, are propelling the transition towards voice-centric interfaces. 
These interfaces expand the capabilities smartphones to devices such as smartwatches, AR-VR headsets, and desktops, thereby becoming more integral to everyday activities. 
\uist{While visual-manual methods such as tapping on smartphones or mouse-clicking on desktops are preferred for their speed and accuracy, \tool leverages the power of LLMs to enhance the intuitiveness of voice interactions. This enables more intelligent mapping of user intentions to visual elements, facilitating the shift to voice-assisted interactions and promoting wider adoption of voice-centric interfaces. Additionally, the features introduced by \tool contribute to the domain of voice-centric research on other devices. For example, breaking down one task into a sequence of actions provides more flexibility than fixed intents (e.g., Firefox Voice~\cite{firefox2021chi}). The continuous learning from historical usage helps these systems understand user commands better and personalize efficient experiences, such as prefilling patient information for Talk2Care~\cite{talk2care2024imwut} and providing personal shortcuts for Firefox Voice~\cite{firefox2021chi}. Moreover, the implementation of anonymization techniques addresses privacy concerns, particularly for voice-centric interfaces that access the display content from screens.}

\uist{Voice-centric interfaces also improve accessibility for users with disabilities~\cite{zhong_raman_burkhardt_biadsy_bigham_2014}. \tool helps individuals with motor and visual impairments by substituting touch-based interactions with voice commands. For motor impairments, this enables easier task completion on mobile devices without touching the screen. Additionally, \tool provides voice shortcuts for visually impaired users, allowing quicker navigation of familiar screens. For example, users can find specific buttons with commands instead of clicking through each one with Talkback. Natural interaction methods could also benefit individuals who struggle with technology and elderly users. }

Despite the promise, challenges such as the effectiveness of voice recognition in diverse environments still persist. Addressing these will be crucial for the broader adoption of voice-centric interfaces, like smart homes and healthcare. This transition, while challenging, opens new avenues for user interaction and emphasizes the need for continued research in the HCI domain.

\textit{LLMs for task automation on user visual interfaces.}
Research has highlighted the capability of LLMs to provide reasoning based on the UI layout, applying to task automation and testing tools~\cite{feng2023prompting}.
These models show remarkable capabilities in incorporating extensive knowledge concerning prevalent app design principles and recognizing standard mobile interface elements, including the toolbar, navigation drawer, and bottom navigation bar~\cite{liu2023chatting} to enhance proficiency in facilitating precise in-app navigation. Our study highlighted the vital role of spatial information and hierarchical UI representations for LLMs in comprehending semantic connections between diverse UI elements, particularly useful for elements lacking textual information like unlabeled icons or images.  In our user study, when tasked with deleting a message lacking a visible delete button, LLM intelligently suggested initiating the process by pressing the unlabelled icon button at the top right, typically the location of the option button, and then selecting \emph{``delete''} from the ensuing options list. The core of this research lies in the transformation of visual interfaces into textual descriptions that LLMs can process, a critical step for enabling effective task execution based on user inputs. Future research should address the models' limitations in unconventional UI scenarios and focus on expanding their adaptability across varied interface designs and complex user tasks. \uist{As models grow, visual models (e.g., GPT-4v) can process images to further enhance the accuracy of such interactions. However, optimizing the usage of visual models to balance accuracy and efficiency is crucial to compensate for the drop in response time, thereby improving their practical application in real-time scenarios.} Such progress in LLM capabilities is pivotal for advancing user interface automation, leading to more user-friendly and efficient digital experiences.

\textit{Towards responsible AI in software systems.} 
In recent years, the remarkable advancements in LLMs have enabled the seamless integration of AI into various software and systems. However, this integration raises significant concerns, particularly regarding data privacy and security~\cite{sun2023does}. The very nature of AI-integrated systems requires access to data, potentially putting sensitive or confidential information at risk. Put in the context of voice assistants on smartphones, users are sceptical as smartphones contain many personal and sensitive data~\cite{kokolakis2017privacy}. Tools like \tool can read such on-screen data and further process them to LLMs. To mitigate these risks, it is essential to implement several key measures, not only to protect users but also to build trust, thereby fostering greater adoption of AI-based interactive systems. \tool represents a pioneering effort in voice-assistive research by applying personal information anonymization to protect user privacy when using LLMs for logical tasks. Additionally, when executing actions on behalf of users, voice assistants must operate responsibly, ensuring that actions do not adversely affect users. This involves seeking explicit user confirmation for decisions, particularly in scenarios where actions may have significant implications, such as replying to important emails or transferring money. Future work in this field should focus on identifying sensitive actions and prompting user confirmation while maintaining a seamless user experience.

\textit{Limitations.}
The current approach poses several limitations. Firstly, the usage-based execution relies prior usage in the particular application, therefore it is inapplicable to unused apps. To address this challenge, our future work aims to develop a more generalized approach to application usage, categorizing apps by their primary functions. For instance, we could devise a standardized set of steps for searching and playing a song that could be applicable across various music applications, thereby simplifying the process for new and unfamiliar apps. Secondly, while our system shows proficiency on Android smartphones, its effectiveness on other Android-based devices remains untested. As previously indicated, there's potential to extend this voice-centric interface to a broader range of gadgets, including smartwatches and AR-VR head-mounted displays. Although the vocal commands might be processed by LLMs across devices, the user interfaces (UIs) of these devices can vary significantly in their logic and layout. For instance, the streamlined interface of a smartwatch might necessitate more concise output due to its smaller screen, while the immersive environment of an AR-VR device could introduce new interaction paradigms. This diversity in UI design and interaction methods across different devices requires more investigations in future works.

\section{Conclusion}
\label{sec:conclusion}

In this paper, we introduce \tool, an innovative virtual assistant designed to enhance user interactions and performance on smartphones. \tool leveraged advanced prompt engineering techniques to harness the capabilities of LLMs for interpreting user commands and constructing logical reasoning components. \tool further streamlined user interactions by automatically storing previous usages to automate subsequent repetitive tasks. Our experiments demonstrated outstanding command interpretation accuracy and the effectiveness of automated execution based on historical usage. In addition, the user evaluation validated \tool's high usability in real-world tasks by improving user performance and reducing mental stress load, aligning with our design objectives. As an open-source project, \tool paves the way for future enhancements in virtual assistant intuitiveness, contributing to the evolution of human-computer interactions. Further research includes applying our versatile database execution approach across diverse platforms and operating systems, as well as exploring innovative prompt engineering techniques to fine-tune LLMs for various reasoning tasks.

\bibliographystyle{ACM-Reference-Format}
\balance
\bibliography{sample}

\end{document}